\documentclass{ws-procs9x6}

\begin{document}

\title{Evolution of Magnetic Field Twist and Tilt in Active Region NOAA 10930}

\author{B. Ravindra}

\address{Indian Institute of Astrophysics,\\
Koramangala, Bengaluru-560034\\
E-mail: ravindra@iiap.res.in\\
www.iiap.res.in}

\author{P. Venkatakrishnan$^*$ and Sanjiv Kumar Tiwari}

\address{Udaipur Solar Observatory,\\
Dewali, Badi Road, Udaipur-313001\\
$^*$E-mail: pvk@prl.res.in}

\begin{abstract}
Magnetic twist of the active region has been measured over a decade using photospheric 
vector field data, chromospheric H$_\alpha$ data, and coronal loop data. The twist and tilt of 
the active regions have been measured at the photospheric level with the vector magnetic 
field measurements. The active region NOAA 10930 is a highly twisted emerging region. The 
same active region produced several flares and has been extensively observed by Hinode. 
In this paper, we will show the evolution of twist and tilt in this active region leading up 
to the two X-class flares. We find that the twist initially increases with time for
a few days with a simultaneous decrease in the tilt until before the X3.4 class flare on 
December 13, 2006.  The total twist acquired by the active region is larger than
one complete winding before the X3.4 class flare and it decreases in later part of 
observations. The injected helicity into the corona is negative and it is in excess 
of 10$^{43}$~Mx$^{2}$ before the flares. 
\end{abstract}


\bodymatter

\section{Introduction}\label{aba:sec1}
It is generally believed that the magnetic fields on the sun is generated underneath 
the convection zone and they rise due buoyancy into the photosphere.
Active region magnetic fields emerge as $\Omega$ shaped fluxtubes at the 
photosphere. These emerged magnetic fields always exhibit some amount of twist.
Twist is shearing of the field lines about its axis. 
Leka et al.\cite{Leka96} have found that the twisted emerging flux carries a 
current along with it. The twist in the active region can originate
due to several reasons.  In the subphotosphere, the field lines could twist due to
the magnetic forces that are internal to fields. The generated magnetic fields can 
be twisted during the dynamo processes\cite{gilman99}. The generated
field while rising through the convection zone is influenced by the convective buffeting
that can introduce the twist in the field lines\cite{longcope98}. In the
photosphere, the differential rotation of the sun can also introduce the twist, though 
in small amount\cite{DeVore00}. The shearing motion of the magnetic footoints can also 
twist the field lines at the photosphere after the emergence.

The parameter $\alpha$ appear in the force-free field equation 
$\nabla \times B$ = $\alpha B$ is considered as a proxy for the twist\cite{burnette04}. 
Active region twist exhibits a 
hemispheric helicity rule, that is left/right handed twist in the northern and 
southern hemisphere\cite{pevtsov95,rust96}. However,
many researchers now believe that the hemispheric helicity rule has to 
be re-investigated due to inconsistencies reported for different phases of 
solar cycles\cite{hagino05, pevtsov08, tiwari09}.

Tilt is the angle made by the line joining the centroids of two opposite polarity 
with the solar equator. Many researchers considered the tilt as a proxy for
the writhe of a fluxtube. The writhe is the spatial deformation of the flux 
tube around its axis. The tilt of the active region also follows the hemispheric
rule that is positive in the northern and negative in the southern hemisphere,
following the Hale-Nicholson law\cite{hale19}. The tilt also depends on the latitude
and is known as Joy's law\cite{zirin88}. The twist and writhe is interchangeable 
and their sum is a conserved quantity\cite{calug59}.

The twist parameter is an interesting quantity in emerging active region\cite{pevtsov03}.
Pevtsov et al.\cite{pevtsov03} have observed that the $\alpha$ increases as the 
active region emerges and reaches
a constant value after a day of the flux emergence. As the flux emerges into the 
photosphere, the coronal helicity increases. The accumulated helicity in the corona 
plays an important role in the energetic events such as flares and coronal mass 
ejections. Nandy et al.\cite{nandy03, hahn05} have studied the relaxation of the twist 
towards constant $\alpha$
force-free fields after flares.  Nindos and Andrews\cite{nindos04}
relate the coronal magnetic field line twist to eruptive and non-eruptive events.

The active region NOAA 10930 is well studied for twist\cite{su09}, flares\cite{isobe07} etc.
In this paper,
we would like to show how the twist and tilt of the active region varies over the 
time. Also, we would like to investigate, if there is any changes in the twist during the 
flare or CME. We used $\alpha_{mean}$ as a parameter to study the twist in the
active region. In the next section, we describe the vector magnetogram 
data utilized in this study and in Section 3, we present the results on how
the twist and tilt evolves with time. In Section 4, we present our conclusions.

\section{Observational Data and Its Analysis}
The spectro-polarimeter is one of the back-end instruments in solar optical 
telescope (SOT\cite{tsuneta08, suematsu08, ichimoto08}) on 
board HINODE\cite{kosugi07} satellite that makes spectropolarimetric measurements at a 
spatial sampling of 0.3$^{\prime\prime}$ in fast mapping mode and 0.16$^{\prime\prime}$ in 
normal mode. The Stokes I, Q, U and V spectra in Fe I 6301.5~\AA~and 6302.5~\AA~lines
are obtained. It makes a map of the active region by spatial scanning of the slit positions.
The spatial resolution along the slit direction is 
0.295$^{\prime\prime}$  and in the scanning direction is 0.317$^{\prime\prime}$/pixel. 
The obtained data set consists of total 
36 vector magnetograms and is a mixer of both the modes. The obtained Stokes signals are 
calibrated using the standard solar software pipeline for the spectropolarimetry. The 
complete information on the vector magnetic fields is obtained by inverting the Stokes 
vector using the Milne Eddington inversion\cite{skumanich87, lites90, lites93}.
The ambiguity in the transverse field is resolved based 
on the minimum energy algorithm developed by Metcalf\cite{metcalf94} and Leka\cite{leka09}. 
The transformation of the magnetic field vectors to the heliographic coordinates
has also been done\cite{venkatakrishnan88}. The error in vertical (B$_{z}$) 
and transverse field (B$_{t}$) measurement is 8~G and 23~G, respectively.

Several methods have been developed to measure the twist in the active
region magnetic fields. Almost all methods are based on the force-free
parameter $\alpha$. The $\alpha_{best}$ was introduced by Pevtsov\cite{pevtsov95}. 
This is the value of alpha for which the computed transverse field is best matched
with the observed transverse field for whole
active region and is called as $\alpha_{best}$. The other methods to measure the
overall twist in the active region magnetic fields are (a) average value of $\alpha$
computed for overall active region magnetic field, $\alpha_{mean}$ 
\cite{hagino04} (b) global alpha, $\alpha_{g}$ and (c) peak value of 
$\alpha$ in the active region, $\alpha_{peak}$\cite{leka05}. Apart from these, there is one more
method that has been introduced in the literature known as signed shear angle (SSA)\cite{tiwari09}.

In this paper, we use the method of $\alpha_{mean}$ to measure the twist in the active 
region NOAA 10930. We express the total twist in the emerging bipole region as a product 
of a magnetic coronal loop length (l) and the winding rate (q). If one takes the
coronal loop as a semicircle of length `l' then the circumference of the semicircle is $\pi d$/2.
Where ‘d’ is the distance between the flux weighted centroids of the two poles.
The winding rate ‘q’ is related to the twist parameter as $\alpha_{mean}$/2. Then the 
total twist is the product of ‘lq’ \cite{leamon03, leka05} and 
is given by,
\begin{equation}
T = \frac{\pi d}{2}\frac{\alpha_{mean}}{2}
\end{equation}

The tilt angle is computed by measuring the tilt of the line joining the flux weighted
centroids of positive and negative poles in the counter-clockwise direction with respect to
the equator. The east side of the line joining the equator is zero and the west side is 
180$^{\circ}$ with values increasing in the counter-clockwise direction as a measure of 
tilt in the active region NOAA 10930. The pixels with vertical field strength values larger 
than 100~G is utilized in computing the centroid. This is to avoid
the small scale features and noisy pixels. 

The alpha maps have been obtained by first computing the vertical current density as
\begin{equation}
J_z(x,y) = \frac{1}{\mu}(\frac{dB_{y}}{dx} - \frac{dB_{x}}{dy})
\end{equation}

once the vertical current density is estimated then the vertical component of the twist
parameter is computed through the relation between $\alpha_{z}$, $J_{z}$, and $B_{z}$ as,
\begin{equation}
\alpha_{z}(x,y) = \mu_{0}\frac{J_{z}}{B_{z}}
\end{equation}

where, $B_{z}$ is the vertical component of the magnetic field in heliographic co-ordinate
system and $\mu_{0}$ is the permeability of free space and its value is 
4$\pi$$\times$10$^{-3}$~G~mA$^{-1}$. To avoid the pixels with zero $B_{z}$, that
always occur near the neutral line, we have used a threshold of 100~G for $B_{z}$.
The $\alpha_{z}$ map is useful for finding the evolution of twist in space and time.

It is found that the measured twist ($\alpha_{z}$) at the photosphere closely
agrees with the coronal field line twist ($\alpha_{corona}$)\cite{burnette04}. 
In order to compute the 
coronal helicity we followed the results obtained by Berger\cite{berger85} and 
implemented to the observed magnetograms as done by D\`{e}moulin et al. \cite{demoulin02,demoulin07}. The 
relative magnetic helicity, using a constant alpha force free method in the linearized 
form is given by,

\begin{equation}
H_{r} = \alpha\sum_{n_{x}=1}^{N_{x}}\sum_{n_{y}=1}^{N_{y}}\frac{\vert{\tilde{B}^{2}}_{n_{x},n_{y}}\vert}{(k_{x}^{2}+k_{y}^{2})^{3/2}}
\end{equation}

\begin{figure}
\begin{center}
\psfig{file=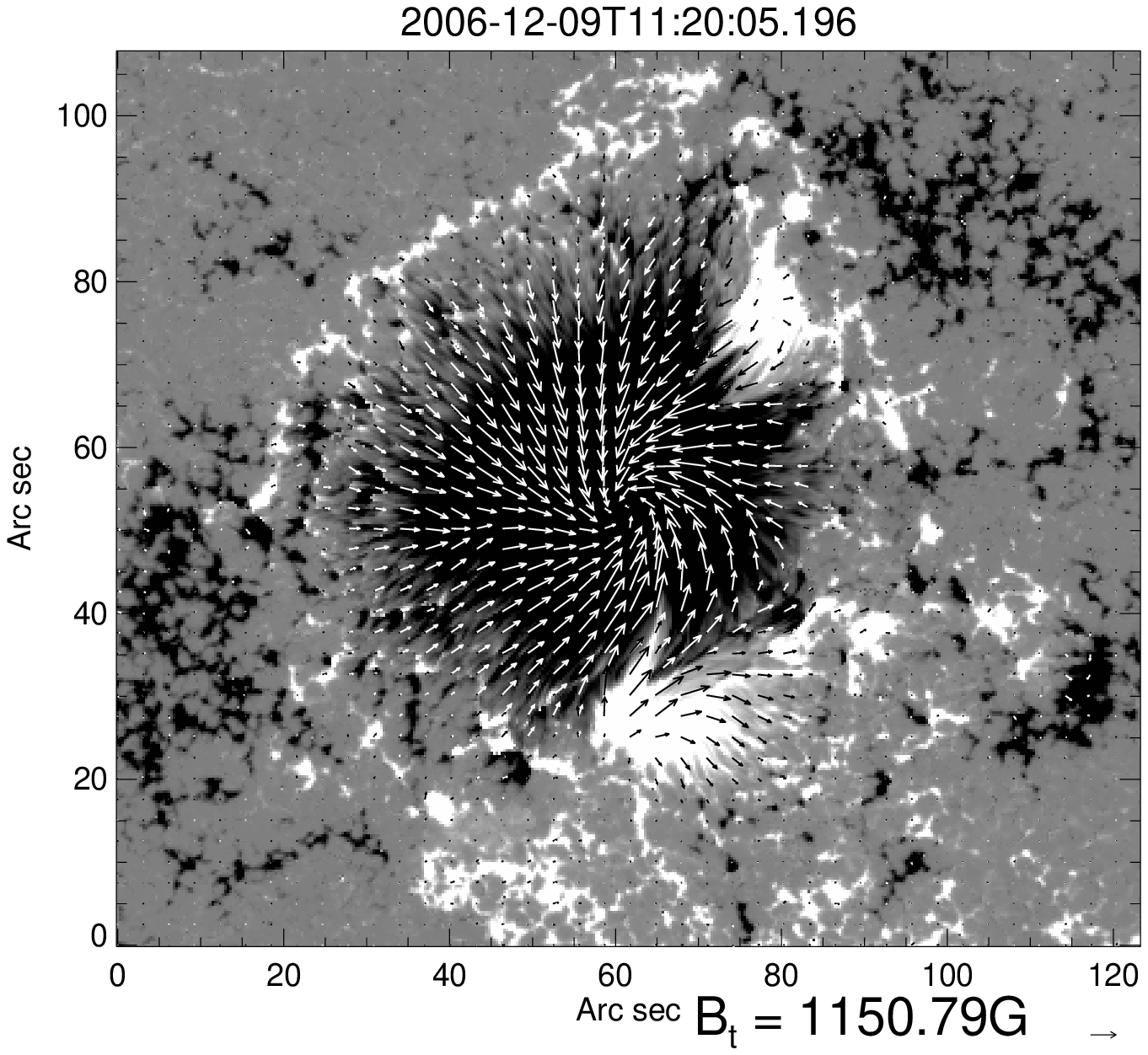,width=2.1in}\psfig{file=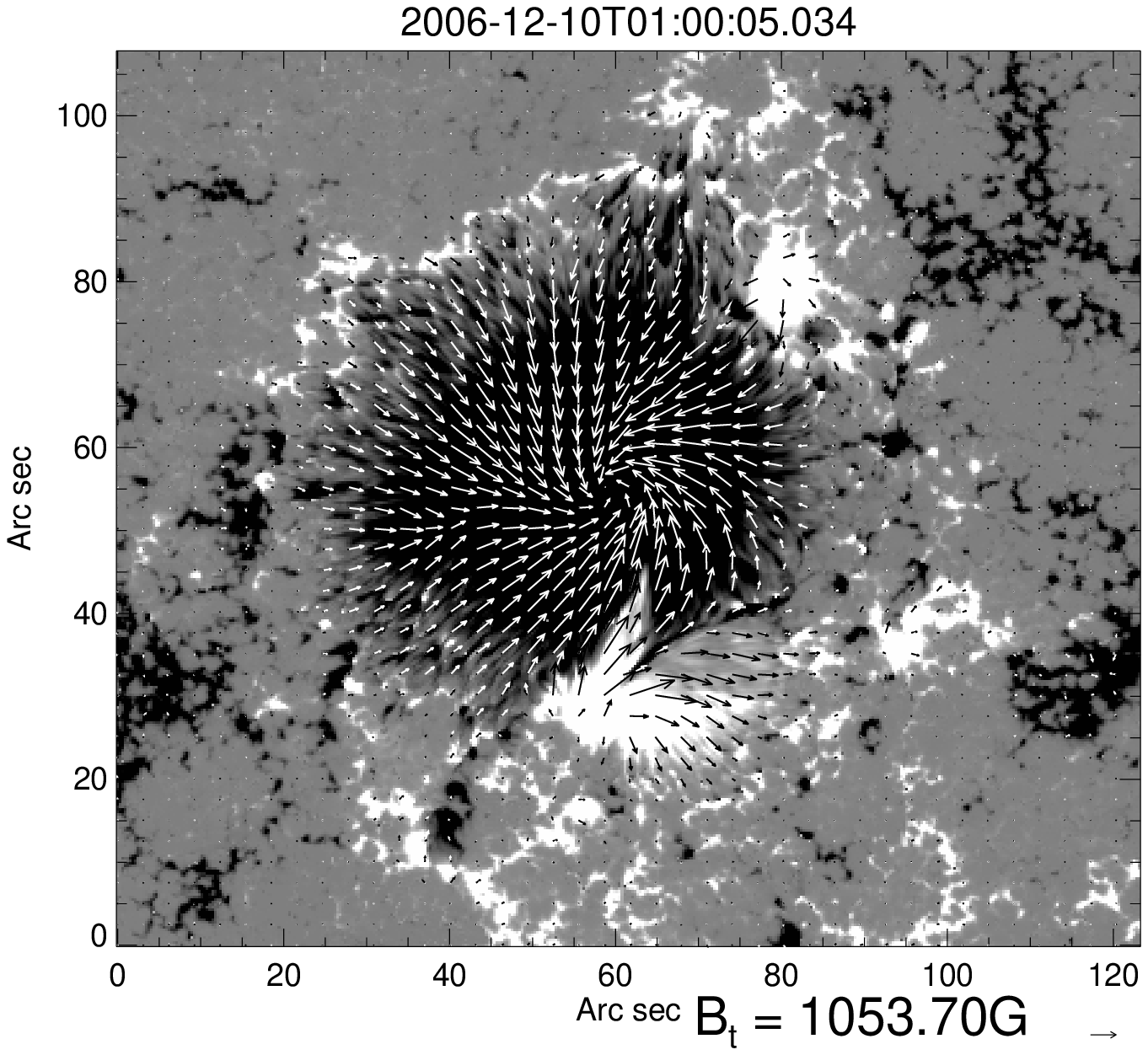,width=2.1in}\psfig{file=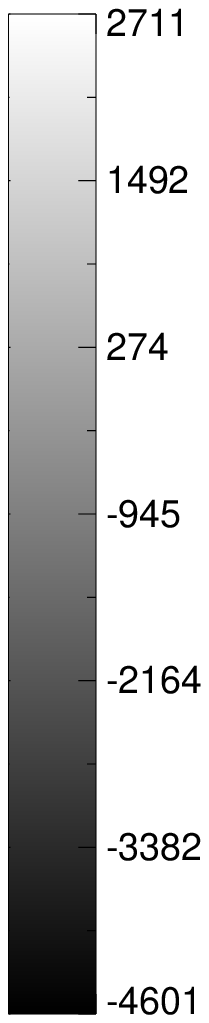,width=0.36in} \\
\psfig{file=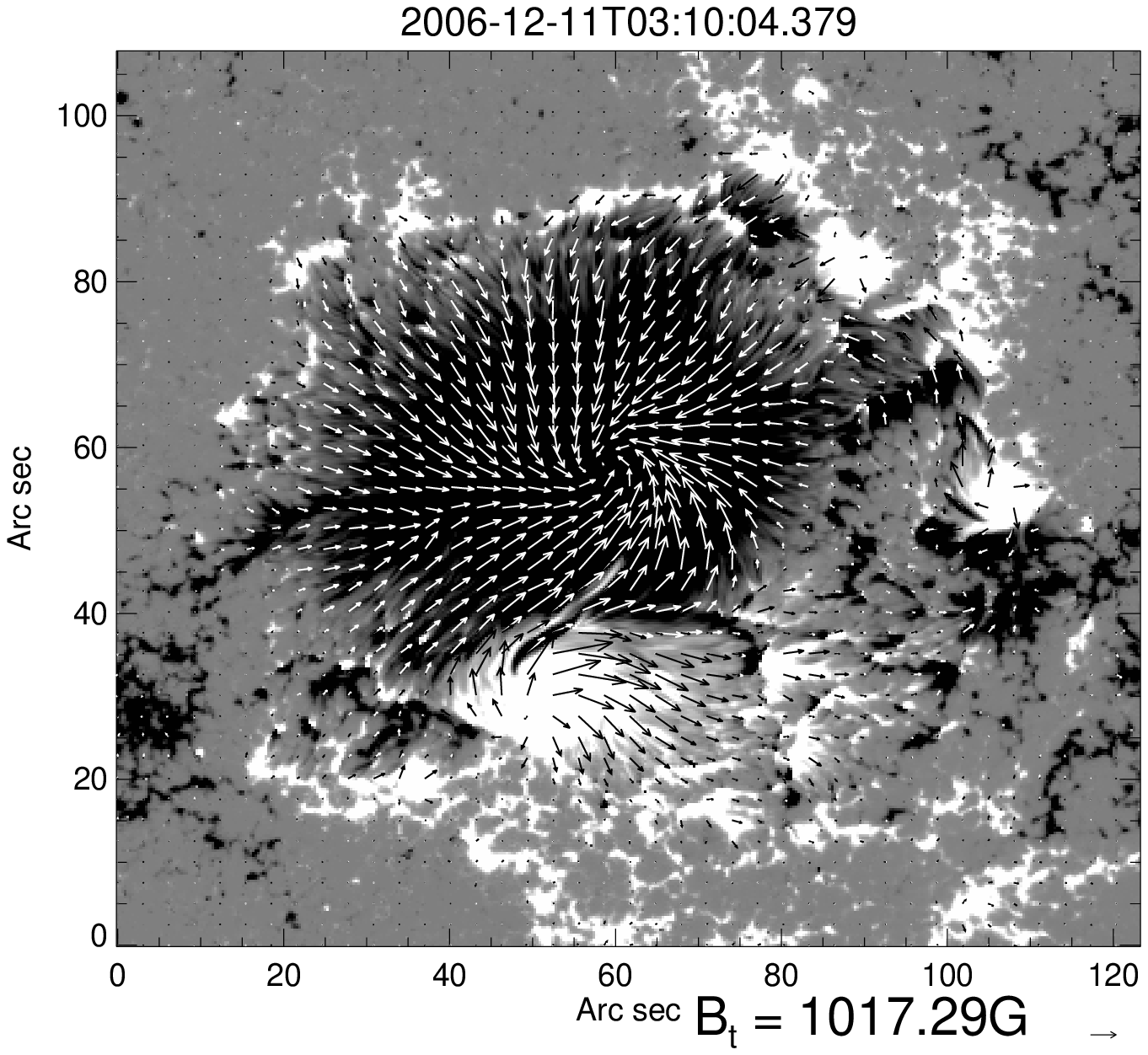,width=2.1in}\psfig{file=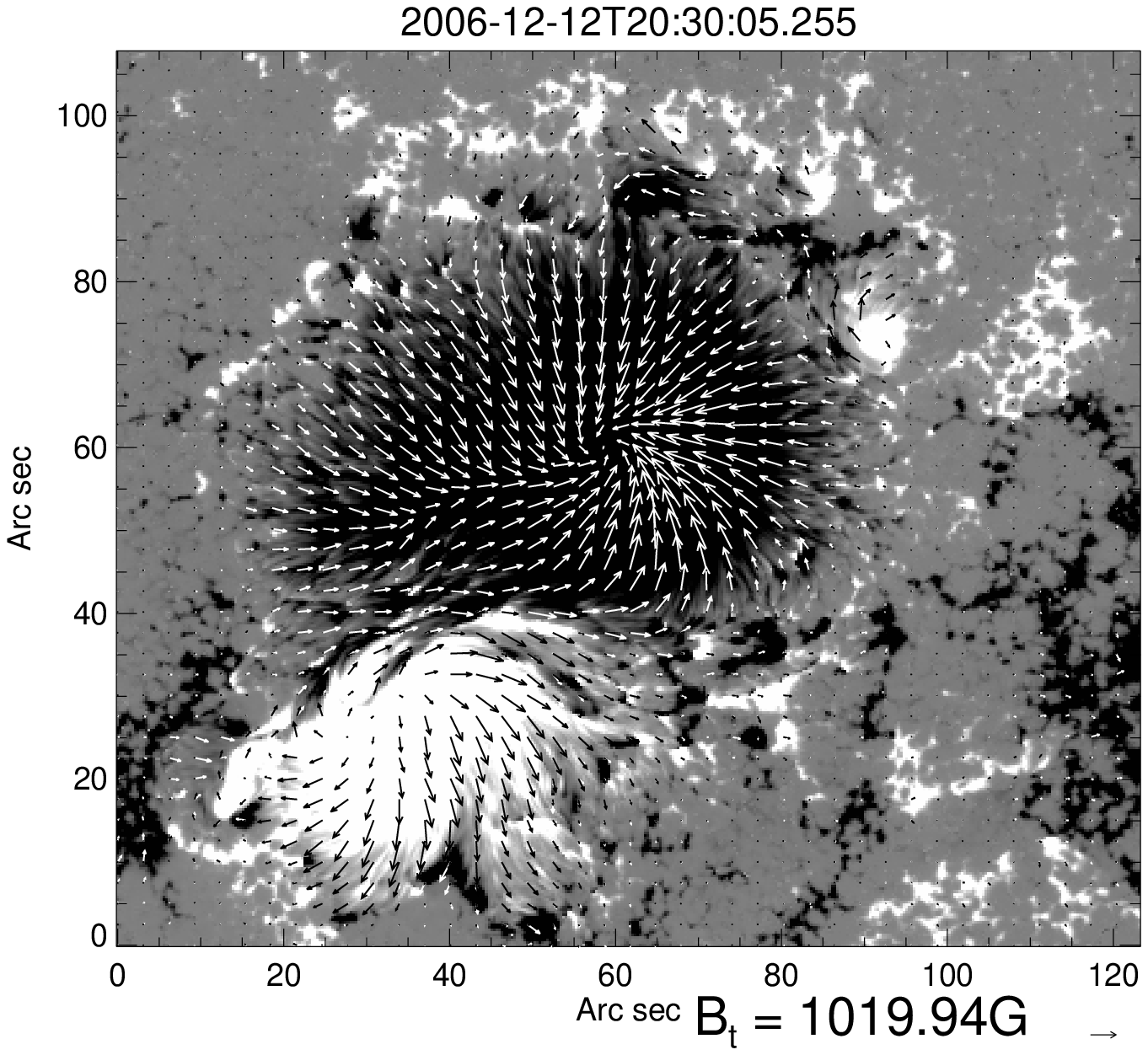,width=2.1in} \\
\psfig{file=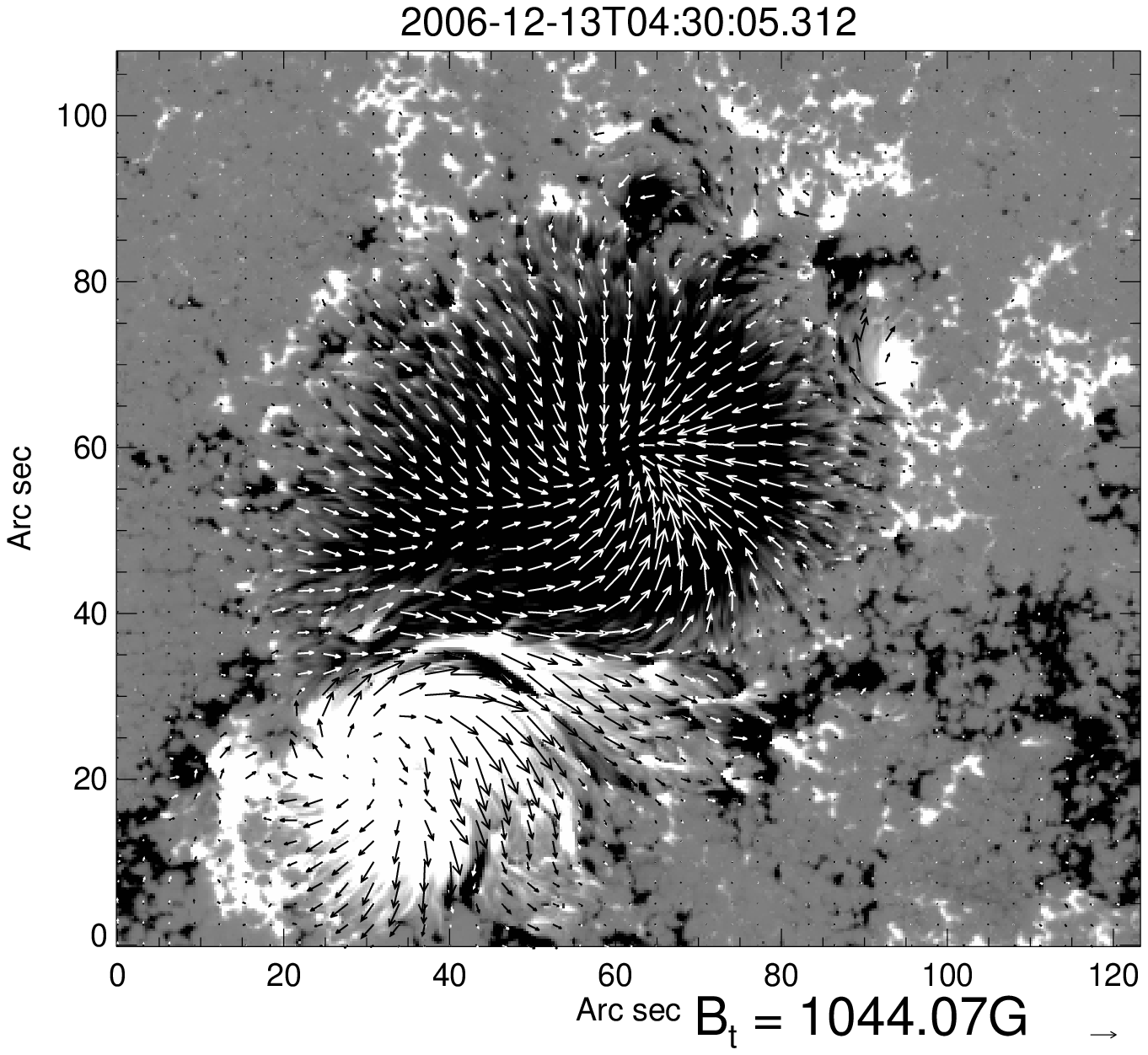,width=2.1in}\psfig{file=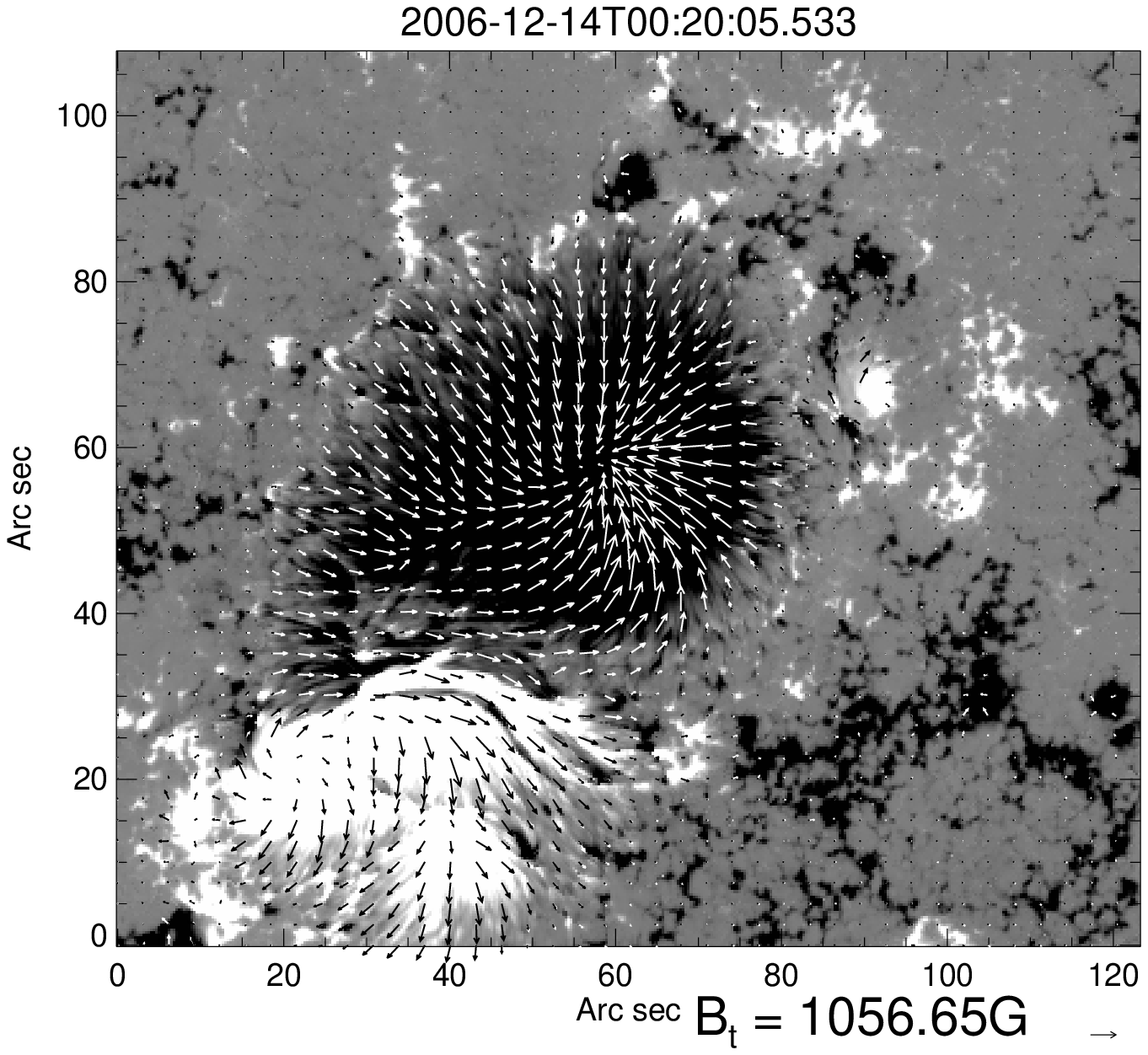,width=2.1in} \\
\end{center}
\caption{The vector magnetograms are shown for each day. The background black (white) 
color represents the S (N) polarity regions of the vertical magnetic field 
strength (B$_{z}$). The arrows represents the transverse field vectors overlaid 
upon B$_{z}$. The size of the arrow at the bottom of each figure indicates the 
magnitude of the transverse field and the vertical bar on the top right figure 
shows the magnitude of B$_{z}$ expressed in G.}
\label{fig:1}
\end{figure}

where, $\tilde{B}_{n_{x},n_{y}}$ is the Fourier amplitude of the longitudinal component
of the magnetic field, $k_{x} = 2\pi n_{x}/L$, $k_{y} =2\pi n_{y}/L$,
and L is the length of the computational window in the horizontal axis.
The original form of the equation, its
limitations and its simplification to the above form can be found in D\`{e}moulin et al. \cite{demoulin02}.
In estimating the coronal helicity we used the $\alpha_{mean}$ as the value for
$\alpha$ in the above equation. 

\section{Results}
The active region NOAA 10930 appeared in the declining phase of the solar cycle 23$^{rd}$ 
in southern hemisphere at a latitude of 5$^{\circ}$. The active region had two polarities
when it appeared on the east limb on December 5, 2006 and they were almost aligned 
perpendicular to the East-West direction. It still survived when it crossed the west 
limb on December 17. 

\subsection{Magnetic Flux}

Figure \ref{fig:1} show the sample images of transverse field 
vectors overlaid upon B$_{z}$ for each day of observations. In the vector field maps 
(as shown in Figure \ref{fig:1}) it is clear that field lines are twisted. Both the 
polarities are 
located close to each other and boundary between the two is highly sheared. The 
N-polarity region was small on December 09, 2006 and it evolved over a period of 7 days.

\begin{figure}
\begin{center}
\psfig{file=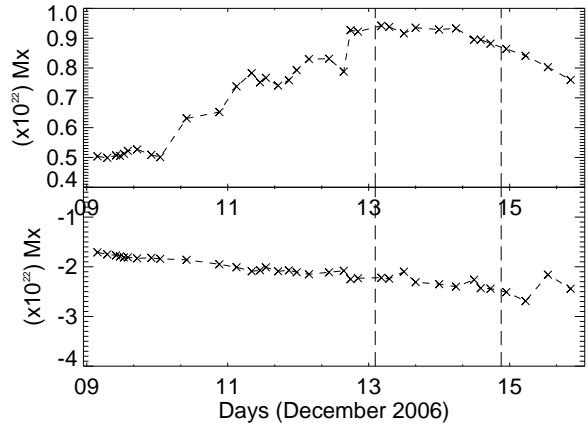,width=3.0in}
\end{center}
\caption{The temporal evolution of magnetic flux is shown as a function of time. 
Top: N-polarity, Bottom: S-polarity.}
\label{fig:2}
\end{figure}

Figure \ref{fig:2} shows the temporal evolution of flux over a period of 7~days. From
the plot it is clear that there is a large imbalance in flux in the N and S-polarities.
A vertical dashed lines in the plot represents the time of X3.4 and X1.5~class flares that
occurred on Dec 13, 2006 at 02:30~UT and Dec 14, 2006 at 21:07~UT respectively. The 
flux in the N-polarity region continuously increased until the X3.4 class flare 
occurred. It remained almost constant for sometimes and decreased subsequently.  

\subsection{Twist in Active Region}
\begin{figure}
\begin{center}
\psfig{file=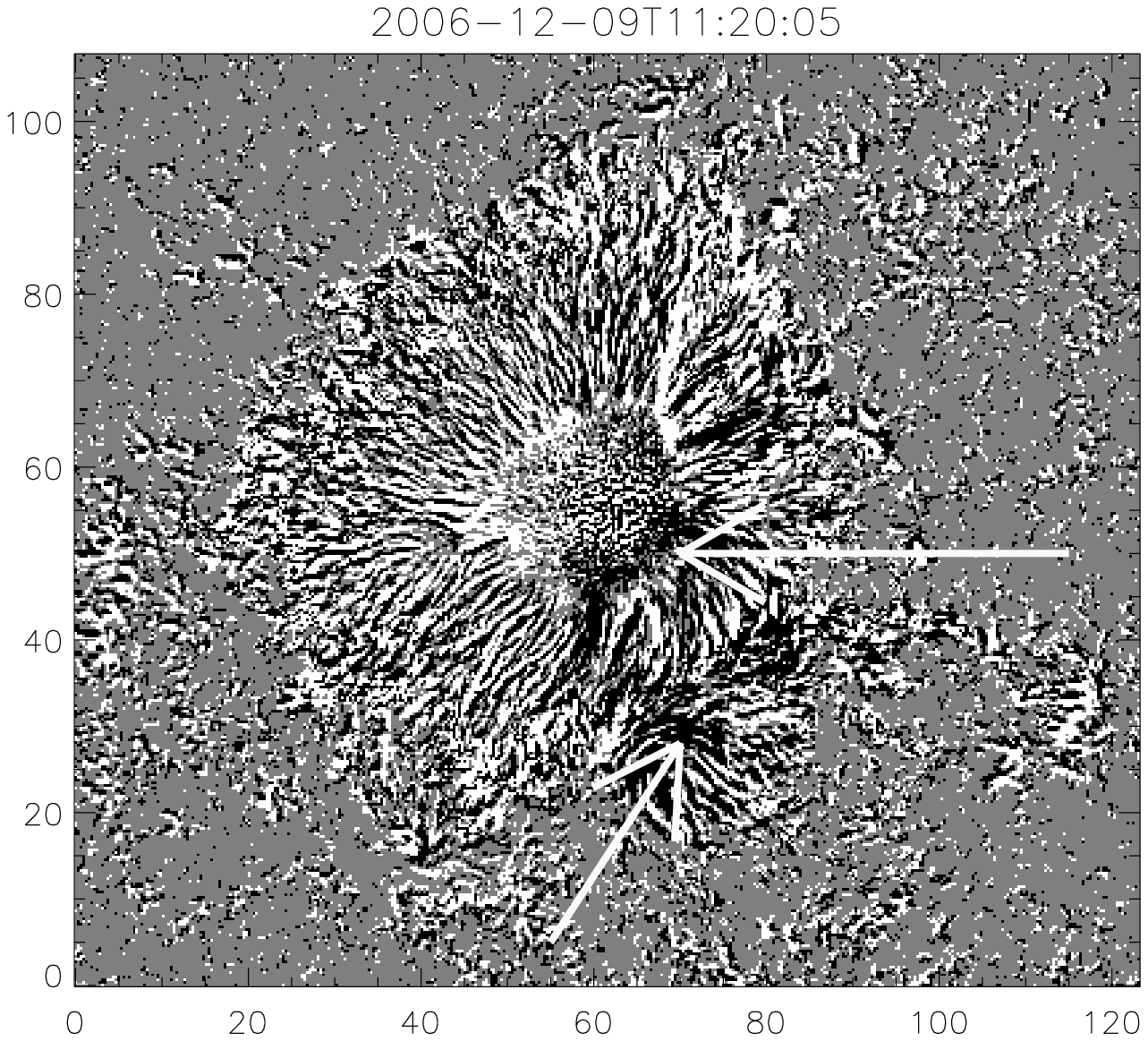,width=2.2in}\psfig{file=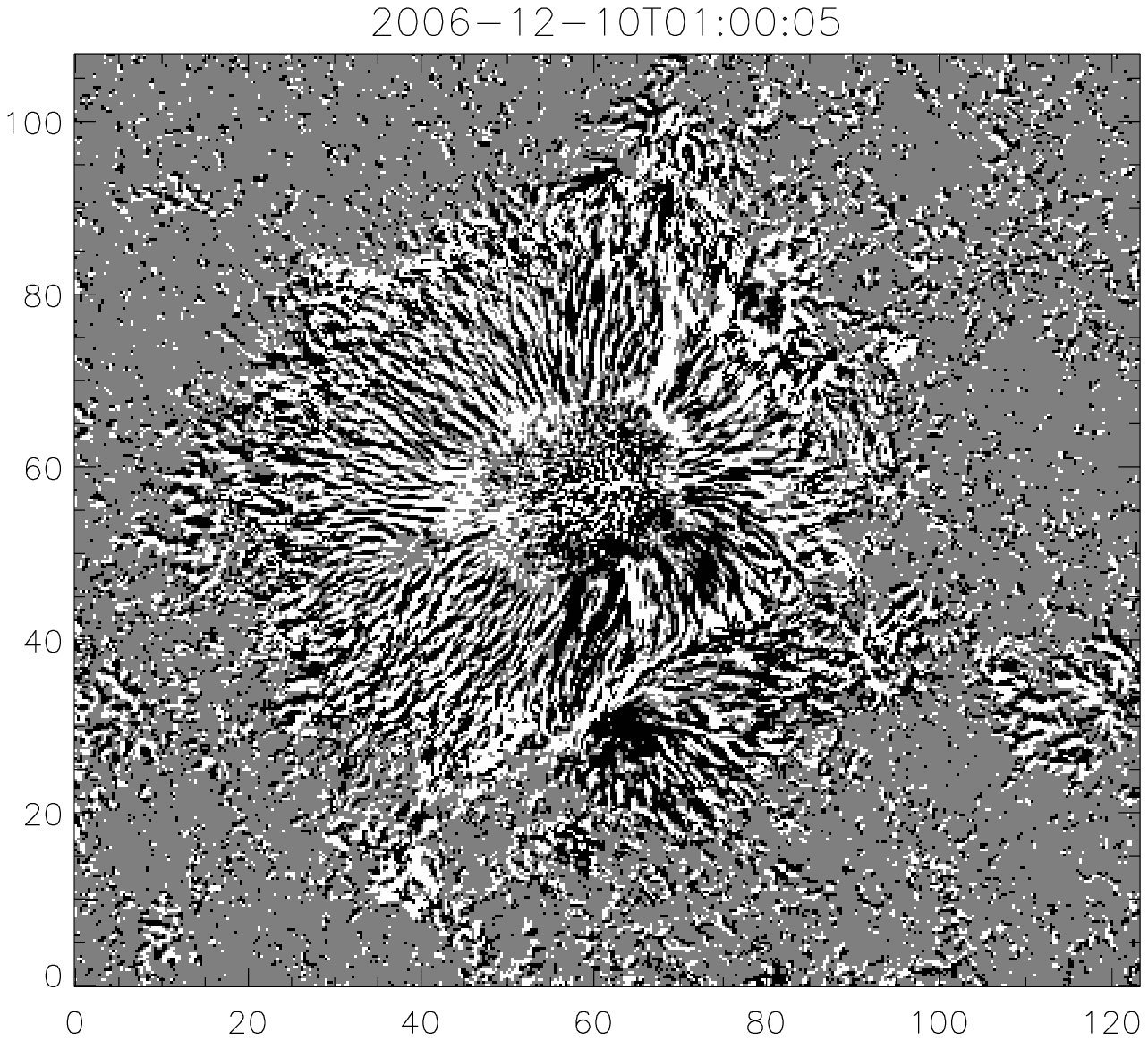,width=2.2in}\psfig{file=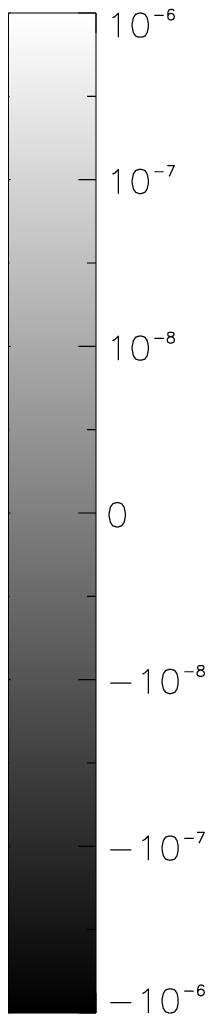,width=0.4in} \\
\psfig{file=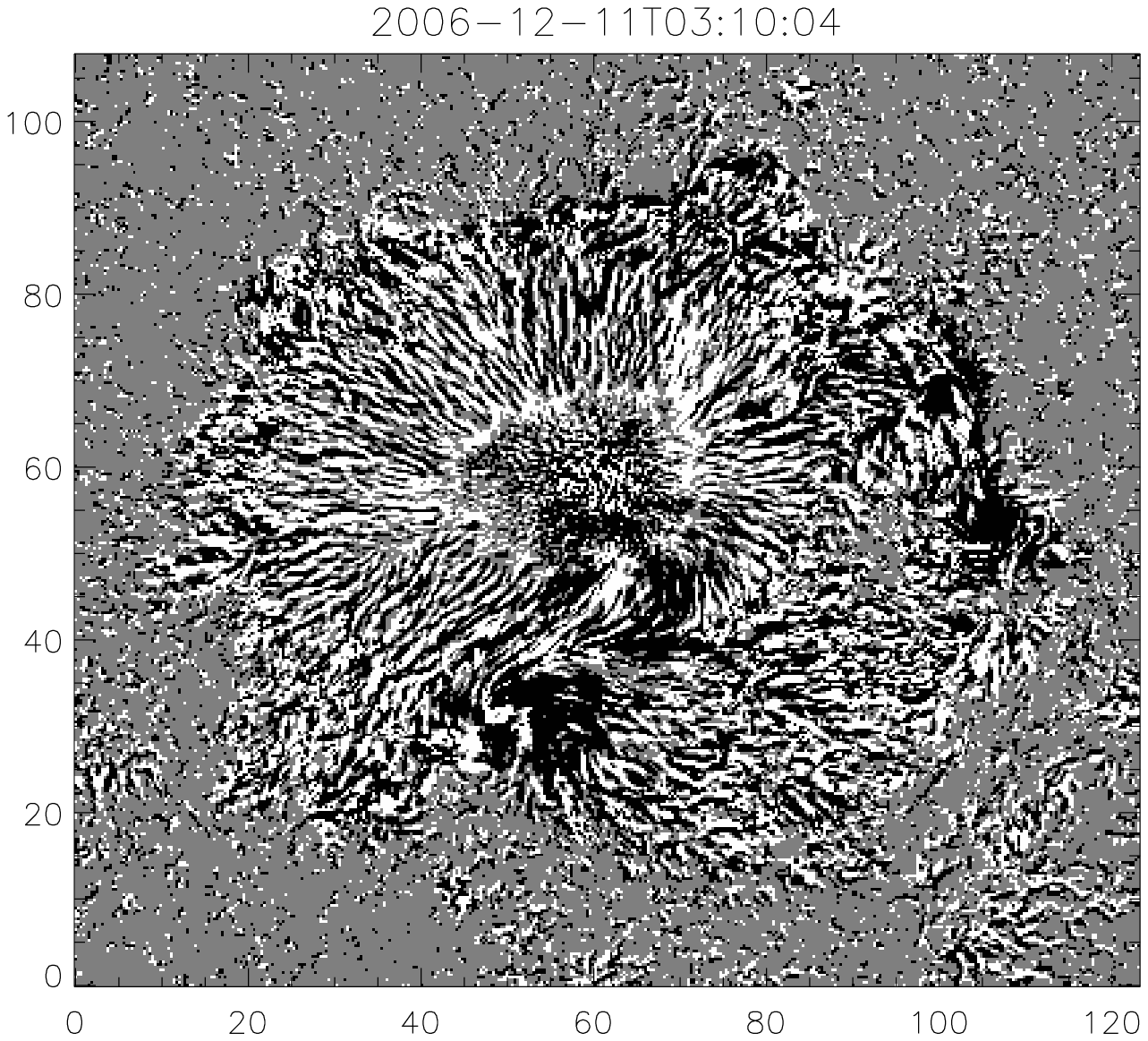,width=2.2in}\psfig{file=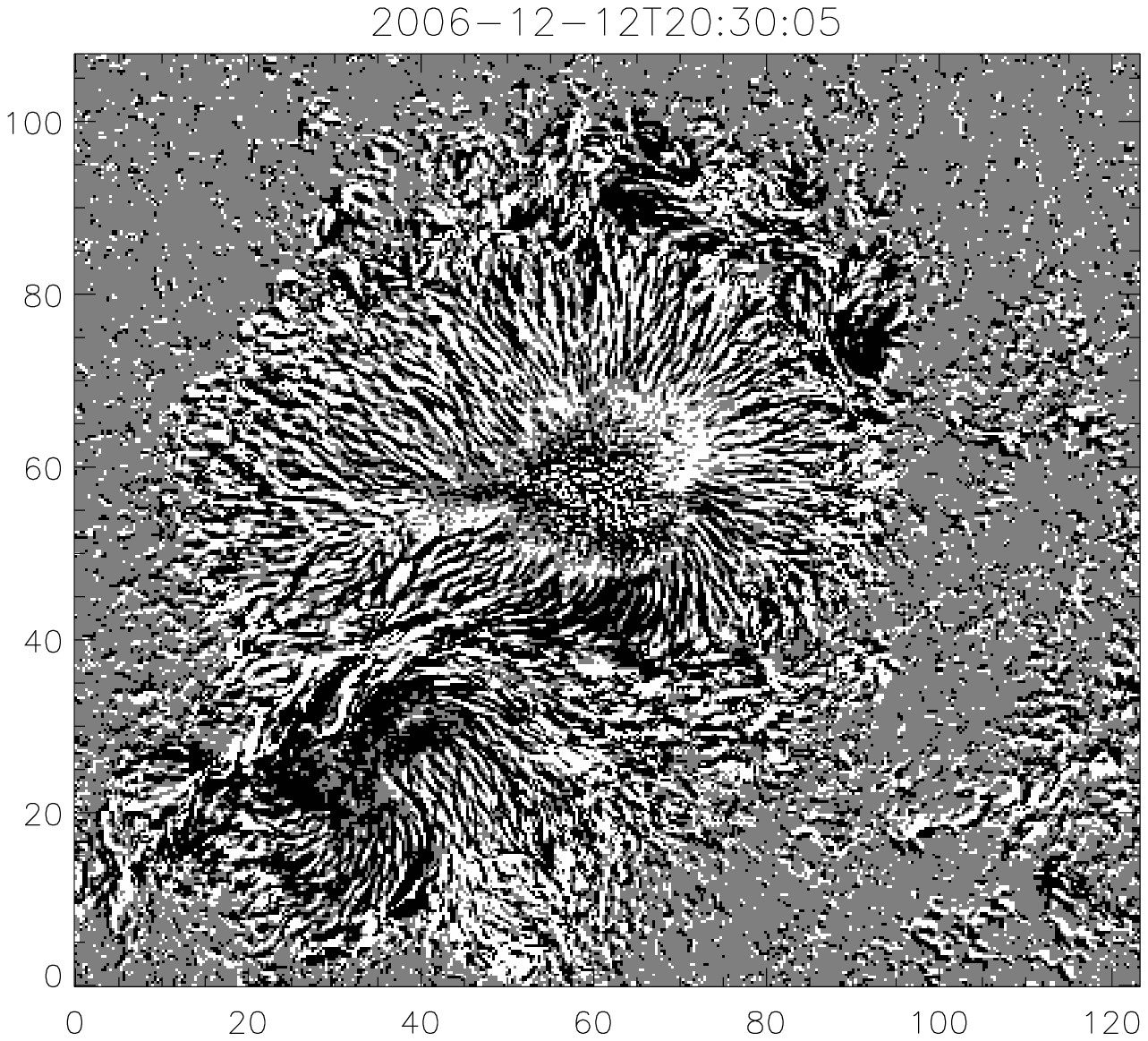,width=2.2in} \\
\psfig{file=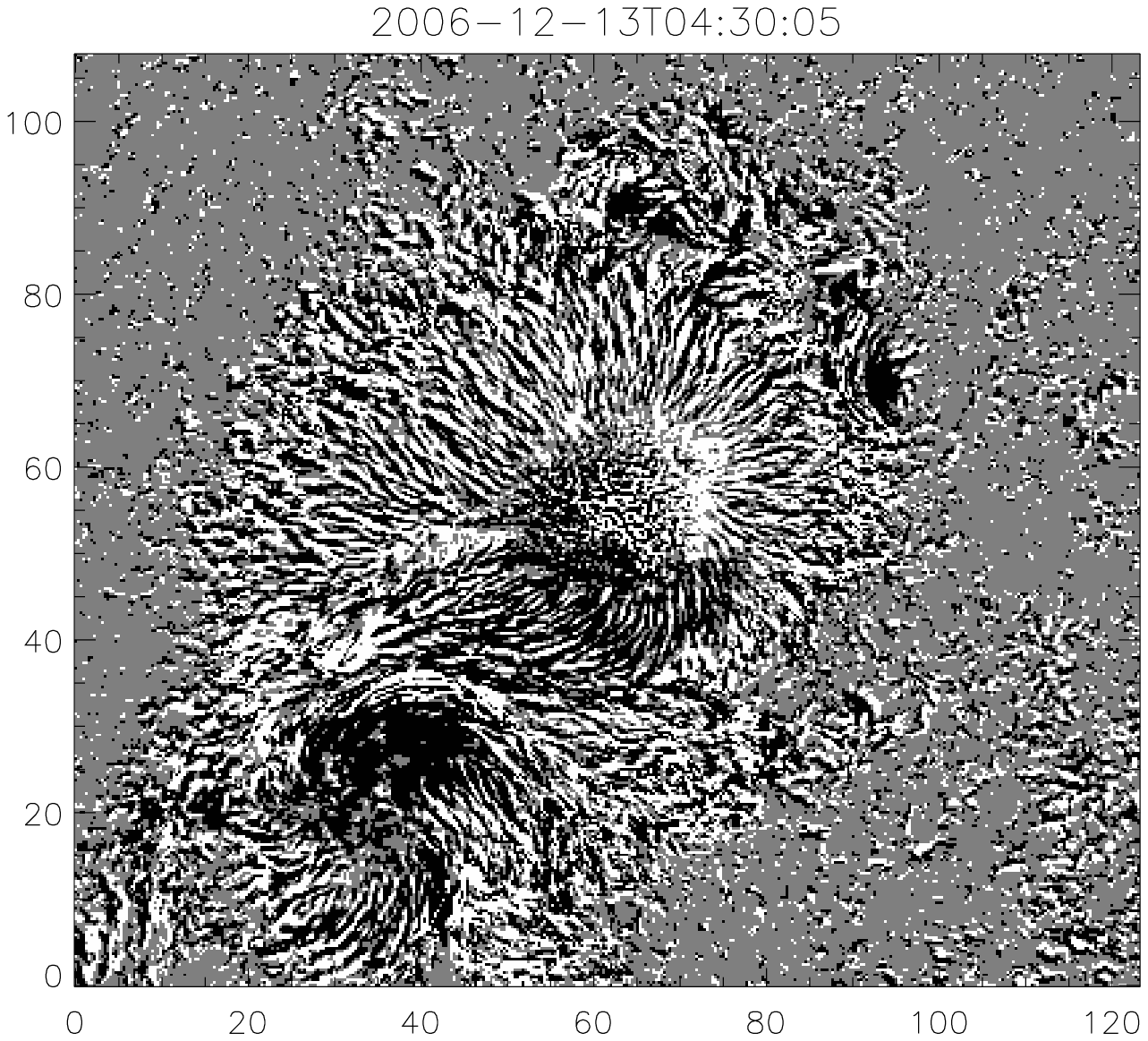,width=2.2in}\psfig{file=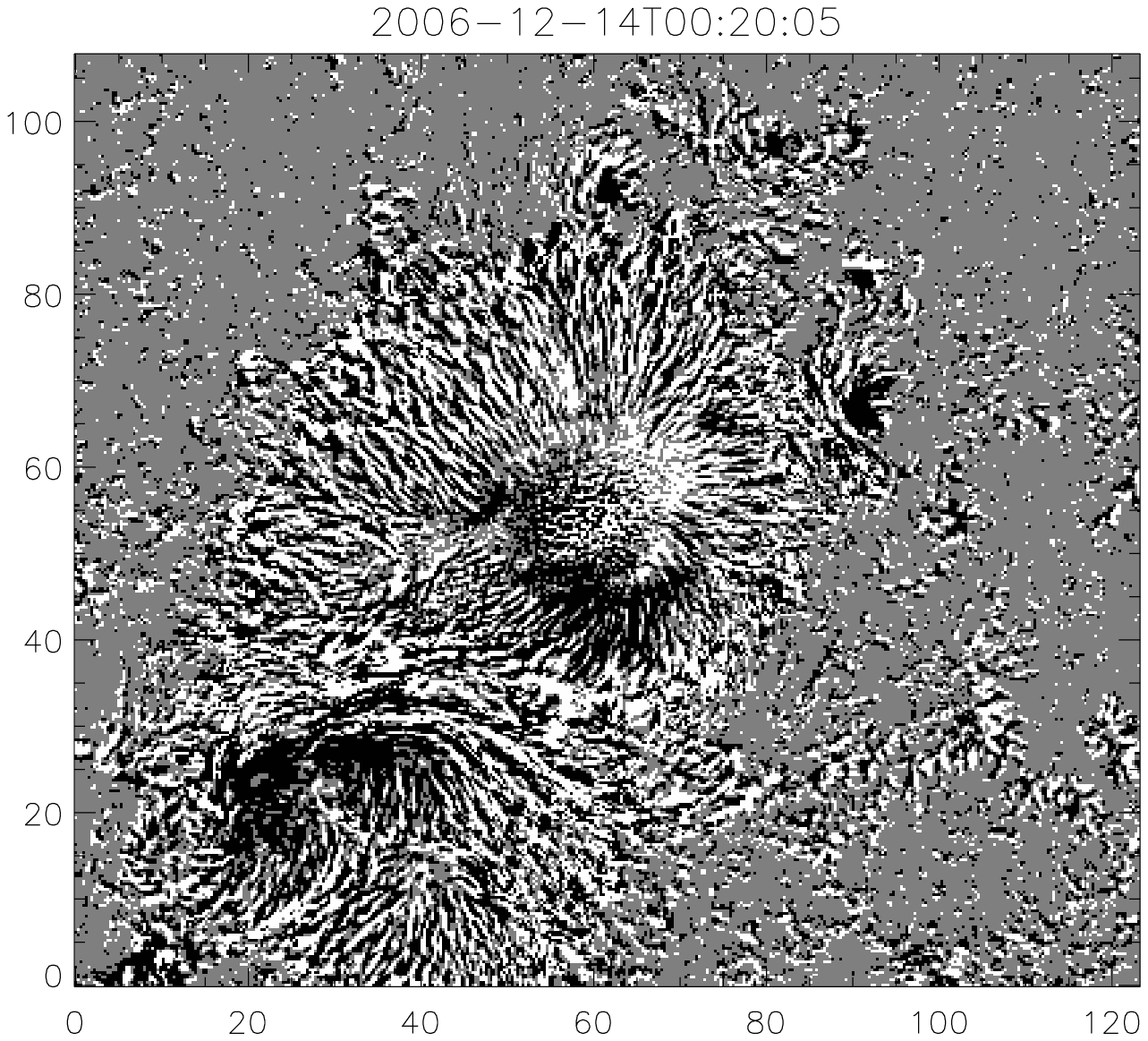,width=2.2in} \\
\end{center}
\caption{The $\alpha$ map is shown for each day of observations. The x-and
 y-axis are in arc sec units. The arrows in the top left image shows the dominant 
negative sign of alpha. The vertical bar on the top right figure shows the range of
alpha in units of $m^{-1}$.}
\label{fig:3}
\end{figure}

Twist is not distributed uniformly over the sunspot. In order to see where exactly
the contribution to the overall twist is coming from, we have made the $\alpha_{z}$ maps.
In $\alpha_{z}$ maps, it is easy to see that there is a mixed $\alpha$ in the 
penumbra as seen by Su et al. \cite{su09, tiwari09}. There is also salt-pepper like 
mixed alpha 
in the umbra. However, a dominant negative sign of $\alpha$ is present in the 
emerging region, the N-polarity sunspot and also in a small portion of S-polarity 
sunspot. These dominant negative sign of $\alpha$ in S and N-polarity region is shown by 
arrows in top-left of Figure \ref{fig:3}. This dominant negative sign of twist can also 
be seen in all the subsequent maps made at different times in both the N and S-polarities.
In the N-polarity region, the area of the dominant negative sign in $\alpha$ increases 
with time till Dec 12 and later it decreases in size.

\begin{figure}
\begin{center}
\psfig{file=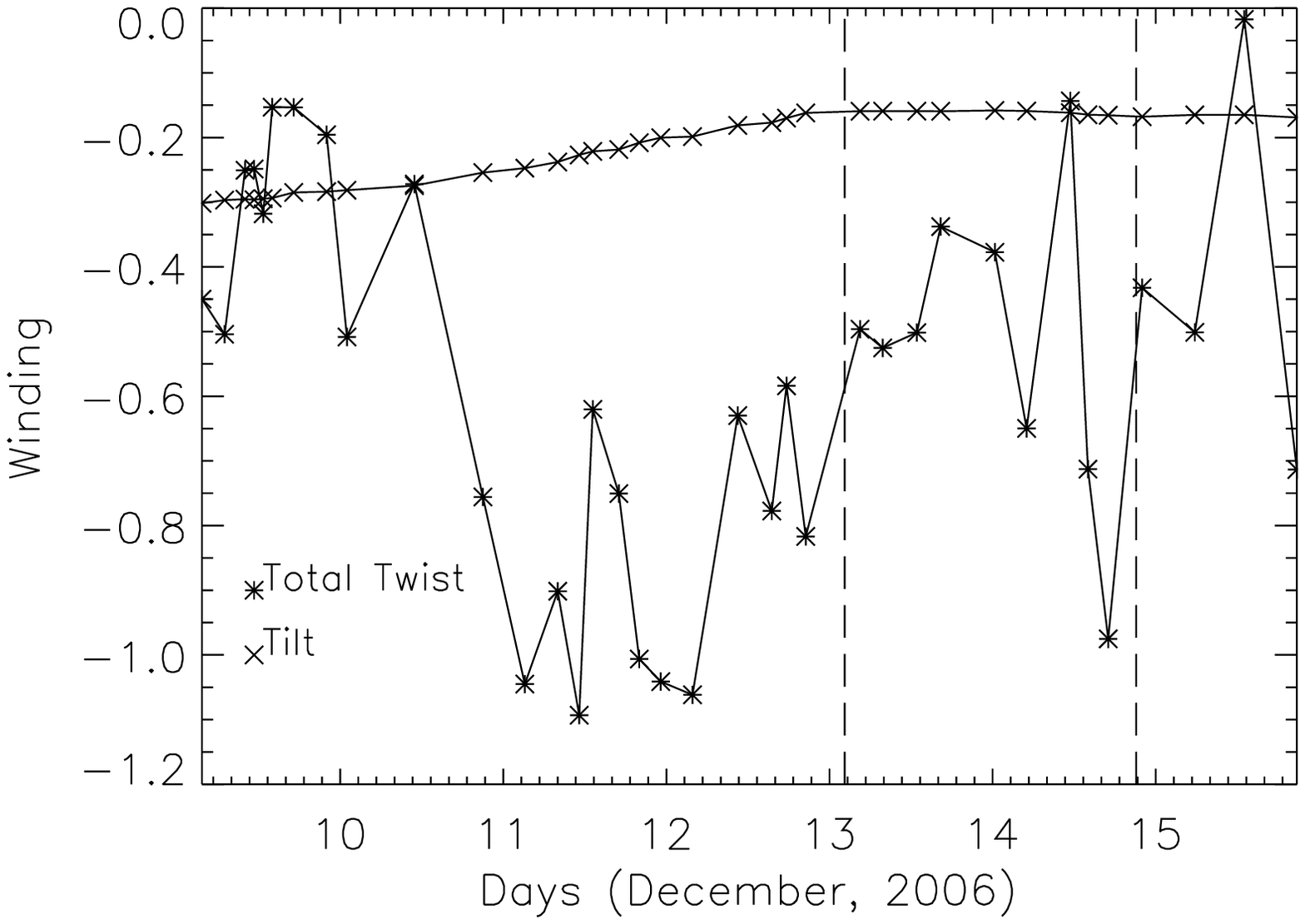,width=2.2in}\psfig{file=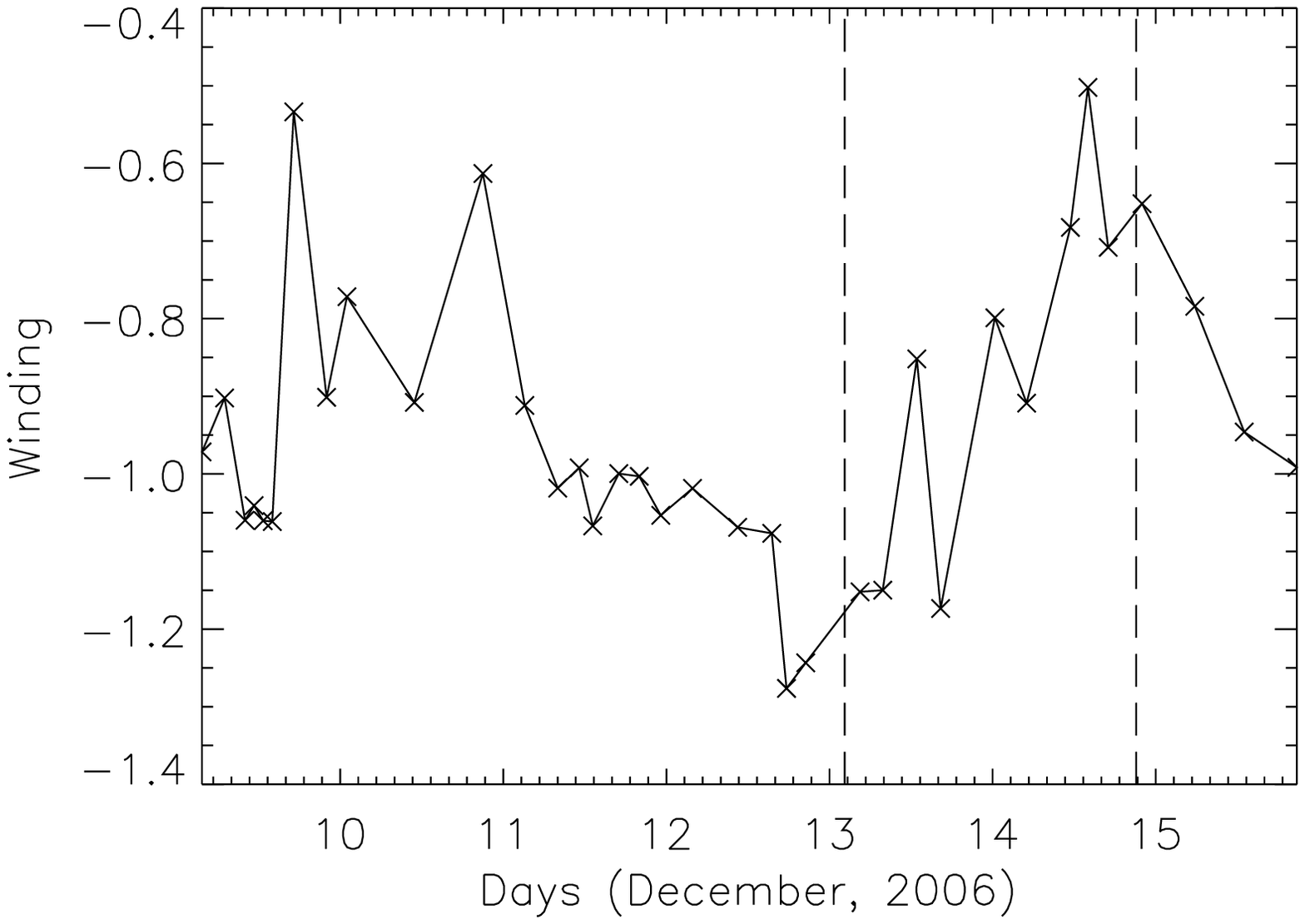,width=2.2in}
\end{center}
\caption{The Total twist and tilt expressed in terms of winding number. Left: For the active 
region as a whole. Right: For the N-polarity region alone.}
\label{fig:4}
\end{figure}

Figure \ref{fig:4}(left) shows the plot of total twist ($\star$) and tilt ($\times$) 
expressed in terms of winding number.
The total twist is estimated using the equation (1), and the number of windings is 
estimated as $T/2\pi$. 
From the plot it is clear that the total twist in the active region increases with time
till the mid of Dec 12, 2006 and after that it starts to decrease with time. The active 
region has got more than one complete winding on Dec 11 which stays there till the 
mid of Dec 12, 2006. The dashed vertical lines in the plot indicate the time of X3.4 and 
X1.5 class flares. Before the X3.4~class flare the active region acquired about 0.8 turns
and it reduced to 0.5 turns of the complete winding after the flare. Similarly, before the 
X1.5 class flare 
the active region has got close to one complete turn, and it decreased to 0.4 after the flare. 
We have estimated the error in measuring the total twist by varying the thresholds in B${z}$
from 50 to 250~G in steps of 50~G. We estimated the mean value of total twist for those
pixels whose value exceeds the thresholds in B$_{z}$. With this method we estimated the 
error in measuring the total twist and it is found to be less than 5\%. This small variations
in total alpha will not affect our results. Apart from these, we also examine the 
temporal variations in total twist  using the another method called 
$\alpha_{g}$ and found a similar variations in the total twist as a function of time. 

The $\times$ symbol in Figure \ref{fig:4}(left) indicates the tilt of the active region with respect to the 
E-W direction. It started to decrease from the beginning of the observation till the beginning
of the X3.4~class flare. It attained almost constant value after the X3.4~class
flare until the end of December 15, 2006. The tilt and twist behaved exactly 
opposite from December 9, 2006 till December 12, 2006. However, they differ
in their magnitudes. This behavior is similar to the results obtained by Holder 
et~al\cite{holder04} and Nandy\cite{nandy06}. However, in the present case as the 
total twist increased with time, the tilt decreased.  

It has been observed that the emerging N-polarity was rotating in the anti-clockwise
direction\cite{min09}. The other S-polarity did not display any rotation. 
In order to see how much the twist appeared in the N-polarity we have independently
estimated the twist for the N-region. Figure \ref{fig:4}(right) shows a plot of total winding
of N-polarity over a period of 7 days.
Similar to the total twist of whole active region,
the N-polarity also exhibited the increase in total twist starting from December 9, 2006
till mid day of December 12, 2006. It accumulated more than one complete winding 
just before the X3.4~class flare. The
winding reduced after the X3.4 class flare, however it was still larger than one 
complete winding. Later, the total twist decreased and before the X1.5 class flare it was just 
3/4th of one complete winding. Again, after X1.5 class flare there is small amount of 
decrease in total twist. It should be noted here that this is a mean value of the 
$\alpha_{z}$ taken from distribution of $\alpha_{z}$. This means that there is a value
of $\alpha_{z}$ larger than the mean value and lower than the mean value. It also
indicates that a significant number of pixels show their total twist larger than 
one complete winding.

\subsection{Relative magnetic helicity}
\begin{figure}
\begin{center}
\psfig{file=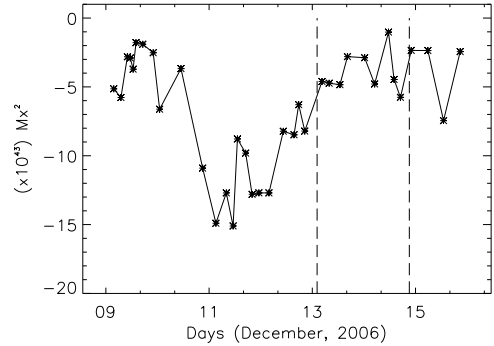,width=2.5in}
\end{center}
\caption{A Plot of evolution of relative magnetic helicity as a function of time is shown
here.}
\label{fig:5}
\end{figure}

Due to the emergence and rotation of sunspots, there is a large amount of helicity
accumulated in the solar corona. There are several ways to estimate the relative magnetic
helicity in the solar corona. Since the cadence of vector magnetic field data is small
we cannot compute the velocity vectors required to estimate the relative helicity
injection. Hence, 
we have used the eq (4) to estimate the coronal magnetic helicity. We have used the 
$\alpha_{mean}$ as $\alpha$ in the eq(4). Figure \ref{fig:5} shows a relative magnetic 
helicity plotted over a period of 7 days. The magnetic helicity curve follows the 
total twist curve. The injected helicity is very large. Before the X3.4 class
flare it was 7$\times$10$^{43}$~Mx$^{2}$ which decreased to 4$\times$10$^{43}$~Mx$^{2}$
after the X3.4 class flare. 

\section{Conclusions}
An increase in the flux and the total twist of the active region during first half 
of the observation suggests that a new flux region is emerging. This confirms 
that the emerging region carries current and twist generated below the photosphere along 
with it as observed by Leka et al.\cite{leka96}. The twist increased in the 
active region until the on-set of X3.4 class flare and showed a decreasing trend 
thereafter. A similar trend of increase and decrease in twist was noticed before and 
after the X1.5 class flare too. This suggests that the twist in the active region
slowly relaxing towards the lower value\cite{nandy03}.

Leamon et al.\cite{leamon03} suggest that the coronal loops become unstable and erupt only
when their total twist exceeds the critical twist, 2$\pi$. However, in their study 
they do not find any active region exceeding this critical twist. Many active regions 
exhibit their total twist of about 2$\pi$/3 and some of them even did not exceed
2$\pi$/6 value of one complete winding. In our 
observations, we find that the total twist exceeded one complete winding on December 11 and
12, 2006. However, it decreased just before the X3.4 class flare reaching
0.8 times one complete winding and before the X1.3 class flare it is close to one 
complete winding. This result was obtained when we take the active region as a whole. 
On the other hand, when we computed the total twist in the N-polarity region alone, 
it exhibited larger than one complete winding and it was about 1.3 times the 1-complete 
winding before the X3.4 class flare. It decreased to 1.1 turn 
after the X3.4 flare. However, before the X1.3 class flare the total twist was less than 
3$\pi$/4 of 1-complete turn. This is only for the mean value of the total twist
in the active region. However, the active region has a distribution of local twist 
($\alpha$) and many regions show larger twist than the mean of this distribution.
This means that this active region shows more than 
one complete winding in many locations of the sunspots, exceeding the critical limit 
of 2$\pi$. We do not know at this stage whether this active region erupted due to
kink instability or due to some other reason.
We need a coronal observations to find such kink instability.
At present, from this study it appears that there is a decrease in the total 
twist after the flare. In order to firmly establish the role of critical twist in
the eruptive X and M class flares, we need to observe many such events.

\section{Acknowledgments}
We thank the two unknown referees for their useful comments which improved the presentation 
in the manuscript.
Hinode is a Japanese mission developed and launched by ISAS/JAXA, with NAOJ as
domestic partner and NASA and STFC (UK) as international partners. It is operated by
these agencies in co-operation with ESA and the NSC (Norway).


\end{document}